\documentclass[11pt]{article}
\usepackage{a4}
\usepackage{epsfig}
\newcommand{\be}{\begin{equation}}
\newcommand{\ee}{\end{equation}}
\newcommand{\bea}{\begin{eqnarray}}
\newcommand{\eea}{\end{eqnarray}}

\newcommand{\ov}[2]{{#1\over #2}}

\newcounter{subequation}[equation]

\makeatletter
\expandafter\let\expandafter\reset@font\csname reset@font\endcsname
\newenvironment{subeqnarray}
  {\arraycolsep1pt
    \def\@eqnnum\stepcounter##1{\stepcounter{subequation}{\reset@font\rm
      (\theequation\alph{subequation})}}\eqnarray}%
  {\endeqnarray\stepcounter{equation}}
\makeatother

\begin{document}

\begin{titlepage}
\hbox to\hsize{%

  \vbox{%
        \hbox{MPI-PhT/22-2000}%
        \hbox{\today}%
       }}

\vspace{3 cm}

\begin{center}
\Large{Static Cosmological Solutions of the Einstein-Yang-Mills-Higgs Equations}

\vskip5mm
\large
P.~Breitenlohner$^*$, P.~Forg\'acs$^\dagger$, D.~Maison$^*$

\vspace{3mm}
{\small\sl$^*$
Max-Planck-Institut f\"ur Physik\\
--- Werner Heisenberg Institut ---\\
F\"ohringer Ring 6\\
80805 Munich (Fed. Rep. Germany)\\}
\vspace{3mm}
{\small\sl
$^\dagger$Laboratoire de Math\'emathiques et Physique Th\'eorique\\
CNRS UPRES-A 6083\\
Universit\'e de Tours,
Parc de Grandmont\\
37200 Tours, France}
\end{center}
\vspace{20 mm}
\begingroup \addtolength{\leftskip}{1cm} \addtolength{\rightskip}{1cm}

\begin{center}\bf Abstract\end{center}
\vspace{1mm}\noindent
Numerical evidence is presented for the existence of
a new family of static, globally
regular `cosmological' solutions of the spherically symmetric
Einstein-Yang-Mills-Higgs equations. These solutions are characterized by
two natural numbers ($m\geq 1$,$n\geq 0$), the number of 
nodes of the Yang-Mills and Higgs field respectively.
The corresponding spacetimes are static with spatially compact sections
with 3-sphere topology.
\endgroup
\end{titlepage}

\newpage
There has been considerable progress in the study of the spherically
symmetric Einstein-Yang-Mills (EYM) and Einstein-Yang-Mills-Higgs (EYMH)
equations, stimulated by the discovery of globally regular solutions
of the EYM eqs.\ \cite{BM}, for a review see e.g.\ \cite{review}.
Up to now most of the attention has been focused on asymptotically flat,
smooth particle-like (self-gravitating sphalerons and monopoles) 
and black hole solutions.

In this paper we present a new `cosmological' 
type of globally regular solutions
of the EYMH eqs.\ describing static, spherically symmetric
spatially compact space-times. Analogous solutions (`static
universes') have already been found in EYM theory \cite{VS1} {\sl
in the presence of a cosmological constant} $\Lambda$, (EYMC). More
precisely, numerical evidence has been presented in Ref.\
\cite{VS1} indicating the existence of a discrete family of
solutions indexed by the number $m$ of nodes of the YM field,
for a special set of values of the cosmological constant,
$\{\Lambda(m),\, m=1,\ldots\}$. The $m=1$ solution is particularly
simple as its energy density is constant and it has a simple
analytic form \cite{CJ,Hos}. It appears that the presence of the cosmological 
constant is essential for the very existence of such solutions.
Therefore at first sight it might appear surprising that such
`static universe' solutions also exist in an EYMH theory {\sl
without a cosmological constant}. In fact this should not come
totally unexpected as the self-interaction potential of the
scalar field, $V(\Phi)$, can generate the necessary
energy density to support a spatially compact space-time.
Intuitively $V(\Phi)$ can act as a `dynamical cosmological
constant'. This viewpoint has been particularly stressed by Linde and
Vilenkin \cite{Linde} in their work on `topological inflation'.
 
The EYMH theory has three different mass scales, the Planck mass 
$M_{\rm Pl}=1/\sqrt{G}$ and the masses 
$M_{\rm W}$ and $M_{\rm H}$ of the YM resp.\
Higgs field, giving rise to two dimensionless ratios 
$\alpha=M_{\rm W}/M_{\rm Pl}$ and $\beta=M_{\rm H}/M_{\rm W}$.   
It turns out to be convenient to introduce an `effective cosmological
constant' $\Lambda=\alpha^4\beta^2/4$.
Our ($m=1, n=0,1,2,\ldots$) EYMH solutions bifurcate with the 
EYMC solution of \cite{VS1} for $\alpha=\sqrt{3},\ldots$ at $\Lambda=3/4$
with a vanishing Higgs field. 
Varying $\alpha$ leads to 1-parameter families ($n=0,1,\ldots$)
with $\Lambda$ determined as a (possibly multi-valued) function of $\alpha$. 
Similarly for some discrete values   
$\alpha_{m,n}$ our solutions bifurcate with the higher nodes cosmological EYMC
solutions of Ref.\ \cite{VS1}, when the Higgs field tends to zero.

Following the notation of \cite{BFM}, we write the spherically
symmetric line element as:
\be  \label{metric}
ds^2=e^{2\nu(R)}dt^2-e^{2\lambda(R)}dR^2
  -r^2(R)d\Omega^2\,.
 \ee
The `minimal' spherically symmetric Ansatz for the YM field is
\begin{equation}\label{Ans}
W_\mu^a T_a dx^\mu=
  W(R) (T_1 d\theta+T_2\sin\theta d\varphi) + T_3 \cos\theta
d\varphi\;,
\end{equation}
where $T_a$ denote the generators of $SU(2)$
and for the Higgs field 
\begin{equation}\label{AnsH}
\Phi^a=H(R)n^a\;,
\end{equation}
$n^a$ denoting the unit vector in the radial direction.
The reduced EYMH action can be expressed as
\be\label{action}
S=-\int dR e^{(\nu+\lambda)}
  \Bigl[
  {1\over2}\Bigl(1+e^{-2\lambda}((r')^2
   +\nu'(r^2)'\Bigr)- e^{-2\lambda}r^2V_1-V_2
\Bigr]\;,
\ee
with
\be
V_1={(W')^2\over r^2}+{1\over2}(H')^2\;,
\ee
and
\be
V_2={(1-W^2)^2\over2r^2}+
{\beta^2r^2\over8}(H^2-\alpha^2)^2+W^2H^2\;.
\ee
Varying  the reduced action~(\ref{action})
one obtains the EYMH equations:
\begin{subeqnarray}\label{feq}
 1-e^{-2\lambda}\left(r'^2+
    \nu'(r^2)'\right)+2e^{-2\lambda}r^2V_1-2V_2&=&0\;,\\
  1+e^{-2\lambda}r'^2-
    2e^{-\lambda}\left(e^{-\lambda}rr'\right)'
    -2e^{-2\lambda}r^2V_1-2V_2&=&0\,,\\
  e^{-\lambda}\left(r'e^{-\lambda}\right)'+e^{-\nu-\lambda}
\left(e^{\nu-\lambda}r\nu'\right)'
+e^{-2\lambda}{\partial( r^2V_1)\over\partial r}+
   r{\partial V_2\over\partial r}&=&0\;,\\
  \left(e^{\nu-\lambda} W'\right)'-e^{\nu+\lambda}
  W\left({W^2-1\over r^2}+H^2\right)  &=&0\;,\\
  \left(r^2 e^{\nu-\lambda}H'\right)'-e^{\nu+\lambda}
  H\left(2W^2+{\beta^2r^2\over2}(H^2-\alpha^2)\right)&=&0\;.
\end{subeqnarray}
Introducing the combinations 
\be
N\equiv e^{-\lambda}r'\,,\quad \kappa\equiv
e^{-\lambda}(r'+r\nu')\, , 
\ee 
the field
eqs.~(\ref{feq}a-e) take the form:
\begin{subeqnarray}\label{taueq}
2\kappa N&=&1+N^2+2\dot W^2+r^2{\dot H}^2-2V_2\,,\\
\dot r&=&N\,,\\
\dot N&=&(\kappa-N){N\over r}-2{\dot W^2\over r}-r{\dot H}^2\,,\\
 \dot\kappa&=&[1-\kappa^2+2{\dot W^2\over r^2}-
{\beta^2r^2\over2}(H^2-\alpha^2)^2-  2H^2W^2]/r\,,\\
\ddot W&=&W\left({(W^2-1)\over r^2}+H^2\right)-(\kappa-N){\dot W\over r}\,,\\
\ddot H&=&2H\left({W^2\over r^2}+{\beta^2\over4}(H^2-\alpha^2)\right)-
(\kappa+N){\dot H\over r}\,.
\end{subeqnarray}
where $\dot f:=df/d\sigma=e^{-\lambda}f'$.
 Note that in Eqs.~(\ref{taueq}) the remaining gauge freedom
(i.e.\ diffeomorphisms in the variable $R$) is implicit in the
choice of the independent variable $\sigma$. A convenient choice (also for the
numerical integration) is $e^{\lambda}=1$.

Next we recall that solutions with a regular origin at $r=0$ \cite{BFM}
have the expansion 
\begin{subeqnarray}\label{bcor1}
  W(r)&=&1-br^2+O(r^4)\;,\\
  H(r)&=&ar+O(r^3)\;,\\
  N(r)&=&1-\left(2b^2+{a^2\over2}+{\alpha^4\beta^2\over24}\right)r^2+O(r^4)\;,\\
  \kappa(r)&=&1+\left(2b^2-{a^2\over2}-{\alpha^4\beta^2\over8}\right)r^2+O(r^4)\;,
\end{subeqnarray}
where $a$, $b$ are free parameters. Solutions satisfying the
conditions~(\ref{bcor1}) which also stay globally regular contain
(among others) asymptotically flat self-gravitating monopoles.
Integrating the field equations~(\ref{taueq}) one finds that for
the generic solution, however, $N(\sigma)$ becomes zero at some
finite $\sigma=\sigma_0$ with finite values of all the other dependent 
variables.
As it is immediately seen from Eq.~(\ref{taueq}b) $N(\sigma_0)=0$
implies stationarity of $r$ at $r(\sigma_0)=r_0$. Then in general
$r(\sigma)$ decreases on some interval for $\sigma>\sigma_0$. We refer
to such a point as an `equator'. In the case of the EYM
eqs.\ it has been proven in Ref.\ \cite{BFMcmp} that for all solutions with
an equator $r(\sigma)$ decreases from $r_0$ all the way to $r=0$ where
a curvature singularity develops (`bag of gold'). 
Adding a cosmological constant radically
changes this conclusion and there exists an infinite family of regular
solutions with 3-sphere topology \cite{VS1}.
\begin{figure}
\vbox{
\hbox to\linewidth{\hss%
\epsfig{bbllx=56bp,bblly=184bp,bburx=567bp,bbury=516bp,width=0.49\linewidth,
        file=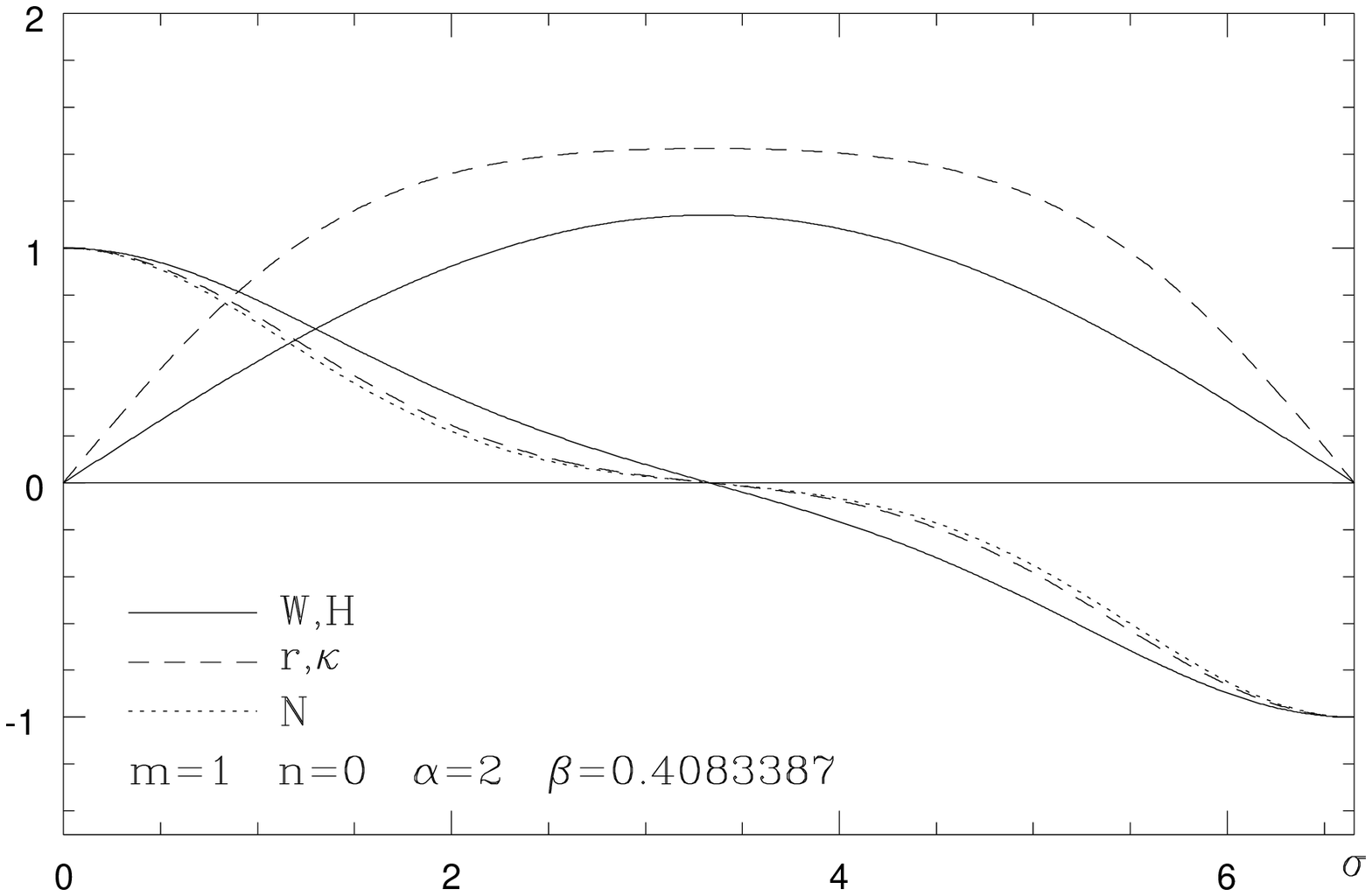}%
\hspace{2mm}%
\epsfig{bbllx=56bp,bblly=184bp,bburx=567bp,bbury=516bp,width=0.49\linewidth,
        file=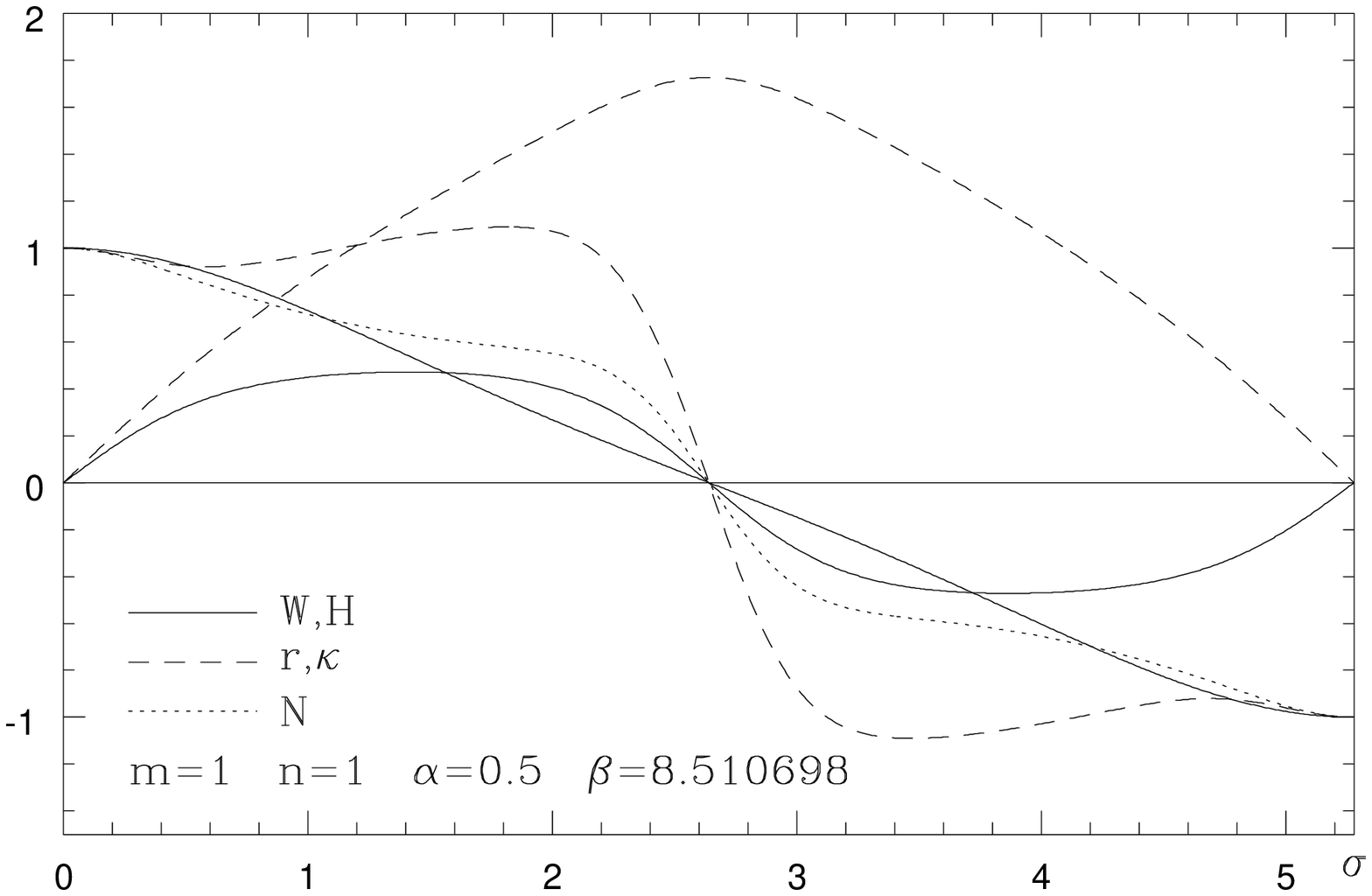}%
\hss}
\vspace{3mm}
\hbox{\hss%
\epsfig{bbllx=56bp,bblly=184bp,bburx=567bp,bbury=516bp,width=0.49\linewidth,
        file=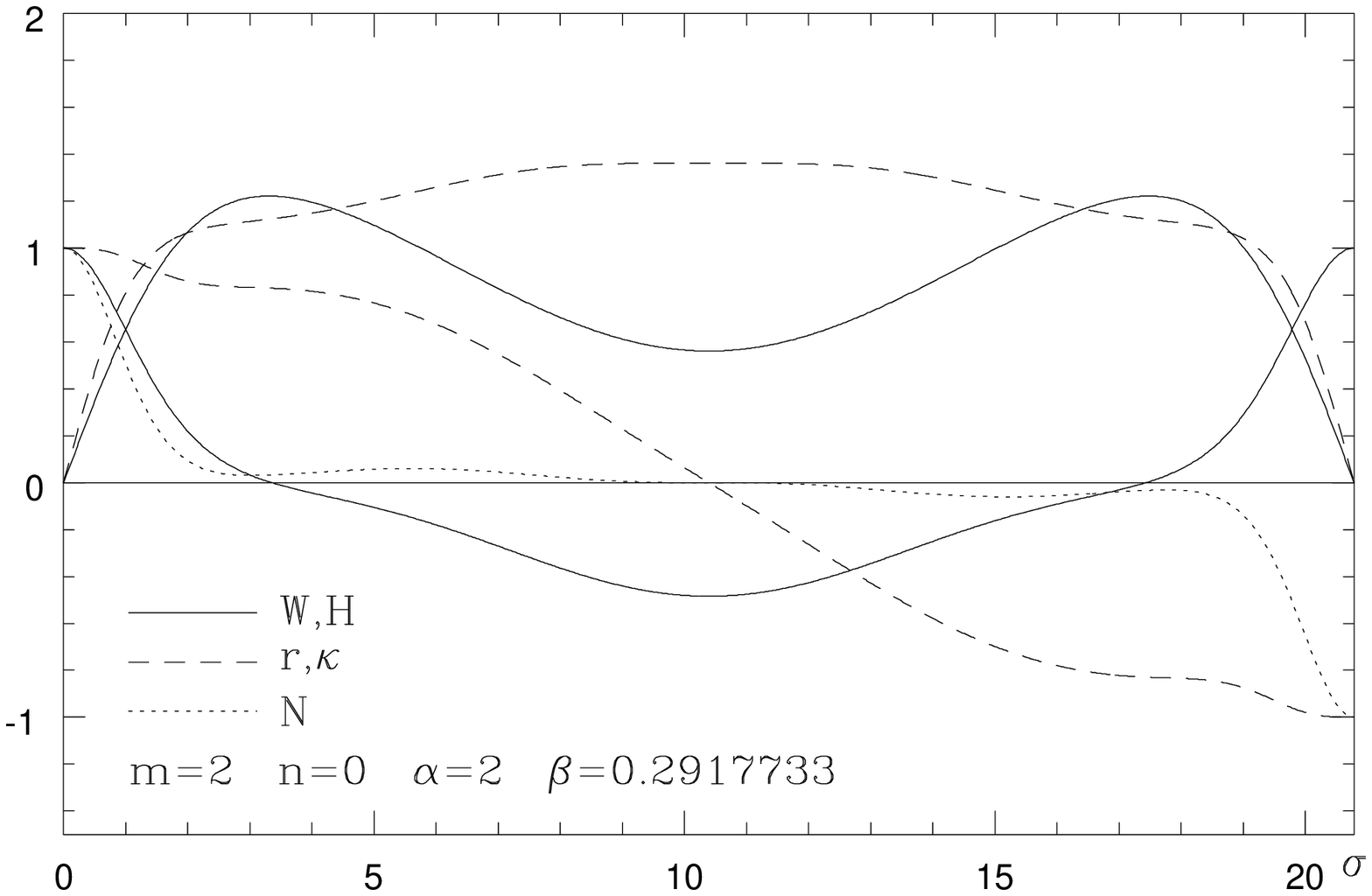}%
\hspace{2mm}%
\epsfig{bbllx=56bp,bblly=184bp,bburx=567bp,bbury=516bp,width=0.49\linewidth,
        file=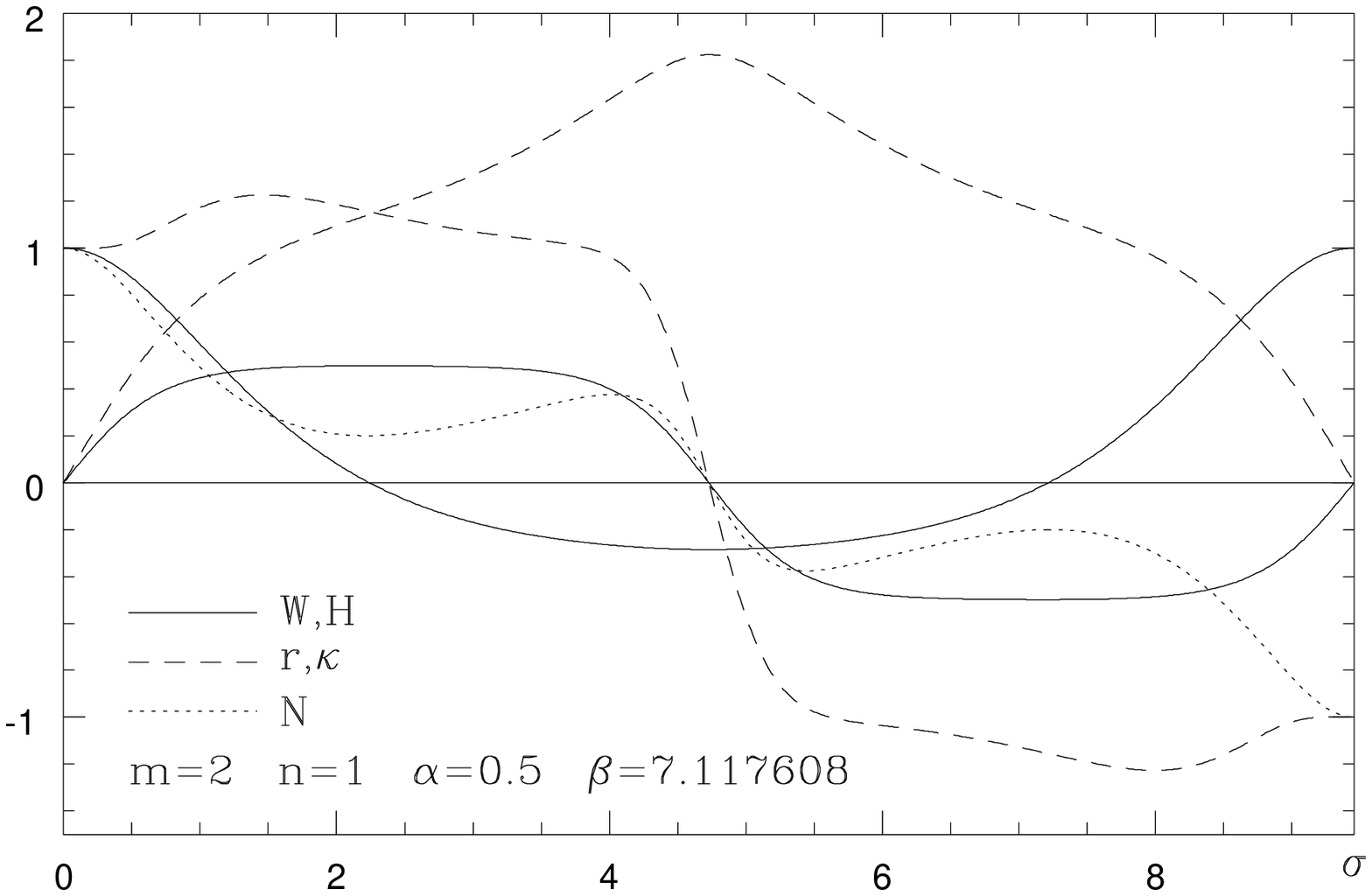}%
\hss}
}
\caption[xpnwm]{Solutions with $m=1,2$ zeros of $W$ and $n=0,1$ zeros of
$H$. The figures display $W$, $H$ (solid), $r$, $\kappa$ (dashed), and $N$
(dotted) as functions of $\sigma$.}
\label{xpnwm}
\end{figure}

Since for $H(\rho)\equiv0$ the EYMH system reduces to an EYMC
theory where the cosmological constant is given as
$\Lambda=\alpha^4\beta^2/4$, it is natural to search for
solutions of Eqs.~(\ref{feq}) with a nontrivial Higgs field
bifurcating with the EYMC ones. To see if such a bifurcating
class really exists one should first establish the existence of
regular solutions of the linearized Higgs-field equation~(\ref{feq}e)
in the background of a 3-sphere type solution of the
EYMC system since the bifurcation occurs for $H\to0$. 
To discuss the linearization of the Higgs-field equation
around solutions of the EYMC equations, it is more convenient to
choose a different gauge from $e^\lambda=1$, namely $\nu=\lambda$ and use the
variable $\rho$ defined by $dr/d\rho=r'=e^\lambda N$.
Then the linearization of the Higgs-field equation~(\ref{feq}e),
around an EYMC background solution ($\Lambda=\Lambda_b$, $r=r_b(\rho)$, 
$\nu=\nu_b(\rho)$, $W=W_b(\rho)$, $H=H_b\equiv 0$) with
$\Lambda_b=\alpha_b^4\beta_b^2/4$) and
$H=h(\rho)$, reads as:
\be\label{lHiggs} \left(r_b^2h'\right)'=2e^{2\nu_b}\left(W_b^2-
{\Lambda_b\over\alpha^2}r_b^2\right)h\,.
 \ee
Let us consider first the simplest, $m=1$
EYMC solution \cite{CJ}:
\begin{subeqnarray}\label{backgr}
W_b&=&\cos x\,,\qquad H_b=0\,, \qquad r_b=\sqrt{2}\sin x\,,\\
N_b=\kappa_b&=&\cos x\,,\qquad
 \nu_b=0\,,\qquad \Lambda_b=3/4\,,
\end{subeqnarray}
where $x=\rho/\sqrt{2}$. Then Eq.~(\ref{lHiggs}) in the
background~(\ref{backgr}), reduces to the following simple
equation: 
\be\label{lineq}
(\sin^2(x)h'(x))'=\left(2\cos^2(x)-{3\over\alpha^2}\sin^2(x)\right)h(x)\,,
\ee 
where now $'$ stands for the derivative with respect to $x$.
We note that Eq.~(\ref{lineq}) can be transformed to a
hypergeometric equation, however, the solutions of interest,
i.e.\ regular on the background geometry and
satisfying the condition~(\ref{bcor1}b), can be directly
found by the following trigonometric polynomial Ansatz for $h(x)$
\be\label{zero} h_n(x)=\sum^n_{k=0}c_k \sin^{2k+1}x\,. 
\ee 
One
then easily obtains the recursion relation for the coefficients
$c_k$:
\be\label{recur}
c_k=\ov{(2k-1)(2k+1)-2-3/\alpha^2}{(2k+2)(2k+1)-2}\,c_{k-1}\,,
\quad k=1\ldots n\,,
\ee
together with the termination condition ($c_{n+1}=0$)
\be\label{eta}
\alpha_{1,n}^2=\ov{3}{(2n+1)(2n+3)-2}\,,\quad n=0,1,\ldots
\ee
yielding $\alpha^2_{1,0}=3$, $\alpha^2_{1,1}=3/13$, 
$\alpha^2_{1,2}=1/11$, \dots
\begin{table}
\caption[alam]{Bifurcation points.}
\label{alam}
\begin{center}\begin{tabular}{rc|ccccc}
$m$&$\Lambda_m$&
$\alpha_{m,0}$&$\alpha_{m,1}$&$\alpha_{m,2}$&$\alpha_{m,3}$&$\alpha_{m,4}$
\\\hline
1&0.75\phantom{00000}&
\phantom{0}1.732051&0.707107&0.480384&0.369274&0.301511
\\
2&0.3642442&
\phantom{0}1.929400&1.080086&0.692265&0.521313&0.420586
\\
3&0.2932176&
\phantom{0}2.694371&1.331930&0.957678&0.712995&0.569524
\\
4&0.2703275&
\phantom{0}3.629807&1.686984&1.170945&0.922451&0.739788
\\
5&0.2608951&
\phantom{0}4.730633&2.061886&1.424943&1.102191&0.912326
\\
10&0.2512791&
12.441074&4.080242&2.777478&2.137747&1.745428
\\
20&0.2501165&
37.982205&8.256034&5.578608&4.284836&3.494772
\\\hline
\end{tabular}\end{center}
\end{table}

That is we have found an infinite family of regular solutions of the linearized
Higgs field equation (bounded `zero modes') in the $m=1$ EYMC background.
These  `zero modes',
indexed by the number of zeros of $h(x)$ ($n=0,1,\ldots$),
indicate the existence of a family
of globally regular solutions of the EYMH equations 
(with $H(x)$ having the same number of zeros) 
bifurcating with the $m=1$ EYMC solution 
for $a=0$ and $\alpha=\alpha_{1,n}$. 

From the $\alpha$-dependence of the r.h.s.\ of Eq.~(\ref{lineq})
it follows that
the solutions $h(x)$ vanishing at $x=0$ have $n+1$ zeros for all
$\alpha\in[\alpha_{1,n+1},\alpha_{1,n})$. According to standard theorems
on Sturm-Liouville operators this number gives also the
number of bound states of the differential operator in Eq.~(\ref{lineq}).
Thus whenever $\alpha$ crosses $\alpha_{1,n}$ from above a new bound state
appears, indicated by the existence of a zero mode for
$\alpha=\alpha_{1,n}$.

\begin{figure}
\hbox to\linewidth{\hss%
\epsfig{bbllx=52bp,bblly=184bp,bburx=568bp,bbury=512bp,width=\linewidth,
        file=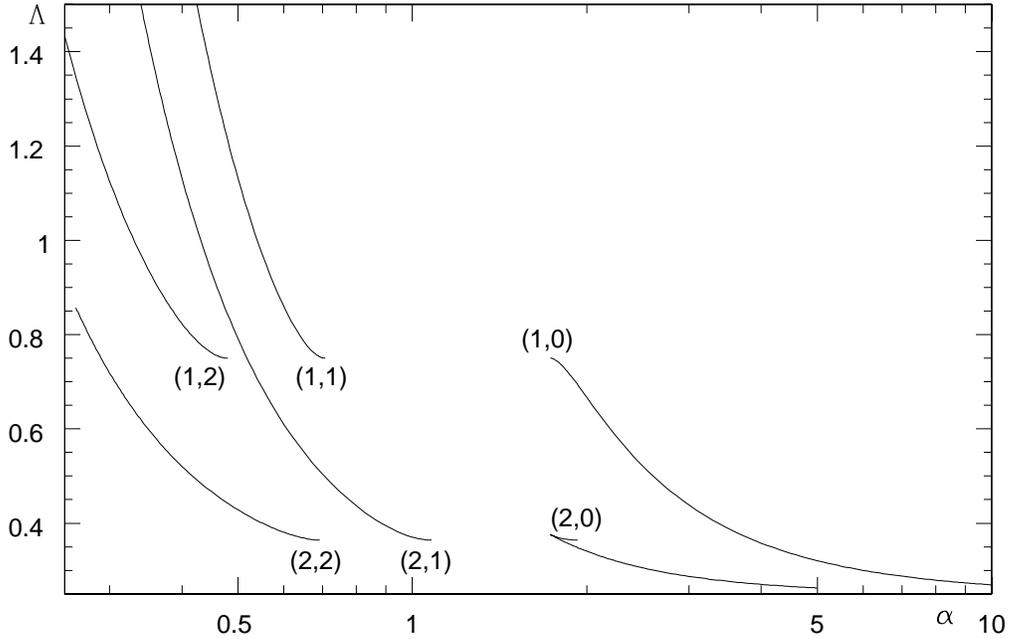}%
\hss}
\caption[xalam]{The effective cosmological constant
$\Lambda=\alpha^4\beta^2/4$ as a function of $\alpha$ for the solutions
with $m,n\leq 2$.}
\label{xalam}
\end{figure}

In contrast to the case $m=1$, the higher node EYMC 
solutions ($m\geq2$) are not known
analytically. Nevertheless one can still conclude 
that for {\sl any} of these EYMC backgrounds an infinite family
of regular zero modes exists. 
To see this, it is sufficient to note that
for $\alpha$ sufficiently small
the r.h.s.\ of Eq.~(\ref{lHiggs}) becomes 
arbitrarily negative over a finite
interval of $\rho$. This implies that the 
number of bound states tends to infinity for $\alpha\to0$.
According to the previous argument the same holds for the
number of zero modes accumulating at $\alpha=0$.
Numerical values for the first five bifurcation points with
$m=1,\ldots,5,10,20$ are given in Tab.~\ref{alam}.

In order to verify the correctness of the above scenario we have
integrated  the field eqs.~(\ref{taueq}) numerically.
To simplify the singular boundary value problem, we have looked only for
solutions, which are (anti)symmetric about the equator. 
This means we impose boundary conditions at $r=0$ and at the equator
(a regular point). In addition to the vanishing of $\kappa$ at the equator,
we require the vanishing of the functions $W$ resp.\ 
$\dot W$ and $H$ resp.\ 
$\dot H$ depending if $m$ and $n$ are even or odd. Thus 
3 functions must have a common zero with $N$ forcing us to tune 3~of the
4~available parameters $\alpha$, $\beta$, $a$, and $b$. 
Fig.~\ref{xpnwm} shows some solutions with $m=1,2$ zeros of $W$
and $n=0,1$ zeros of $H$.

In Fig.~\ref{xalam} we have plotted the values of the parameters $\alpha$ and
$\Lambda$ for the solutions
with $m,n\leq 2$. While $\alpha$ runs to large values with decreasing
$\Lambda$ for the $n=0$ solutions, i.e. with a nodeless Higgs field, the 
parameters for the solutions with $n\geq 1$ show the opposite behaviour. 
This structure remains true for higher values of $m$, although 
the change occurs in general for some $n_0>1$ increasing with $m$ (e.g.\
$n_0=6$ for $m=10$).
Similar to the case $(2,0)$ some of the graphs in the $\alpha$-$\Lambda$
plane for $m>2$ show a minimum of $\alpha$, 
before they tend to large values of $\alpha$ (e.g.\ $n=1,2,4,5$ for $m=10$).

The families of numerical solutions for which $\alpha$ becomes 
large seem to have a
smooth limit for $\alpha\to\infty$ with $\Lambda\to1/4$.
In this limit, assuming that $H$ stays finite,
Eqs.\ (\ref{feq}) reduce to an EYMH system with a cosmological constant 
$\Lambda=1/4$ instead of a Higgs potential ($\beta=0$). 
The spatial sections of the corresponding space-times are no longer compact
($\sigma\to\infty$ and $r_0\to\sqrt{2}$).
We have also numerically integrated the limiting field eqs.\ and
our results fully confirm the existence of the limit $\alpha\to\infty$.

It was argued in \cite{Brodbeck} that the solutions of the EYMC
system of \cite{VS1} are unstable. 
As usual the criterion for instability is the existence of imaginary modes
of the linearized time-dependent field equations.
Although we do not quite approve of the methods of \cite{Brodbeck}, 
we nevertheless believe that their result is correct. For the case $m=1$
the instability has already been shown in \cite{Hos}.
We expect the instability of the YM system to persist in our case with the
Higgs field with the same number of unstable modes at least for small values
of $a$. In view of the discussion of the solutions of
Eq.~(\ref{lHiggs}) given above there are additional instabilities of the
EYMC solutions viewed as solutions of the EYMH theory (with $a=0$) in the
Higgs sector. More precisely, there are $n+1$ unstable modes for
$\alpha\in[\alpha_{m,n+1},\alpha_{m,n})$. 
Turning to solutions with $a\neq 0$ but small the 
only questions is, if the number of the unstable
Higgs modes of the $(m,n)$ solutions is equal to $n$ or $n+1$. 
The answer depends on the behaviour of the zero mode at the bifurcation 
point $\alpha_{m,n}$ when $a$ deviates from zero. If it turns into a bound
state one gets $n+1$, if it moves into the continuum $n$ unstable modes.  

Similar to the globally regular solutions discussed in this paper, 
there are families of solutions with a horizon branching off from the
corresponding solutions of \cite{VS1} without a Higgs field. In fact there
are even more general solutions having two horizons, one of them replacing
the regular origin of the forementioned class. We plan to give a detailed
account of these solutions in a forthcoming publication.

\end{document}